\documentclass{article}
\usepackage{dcase2022,amsmath, amssymb,graphicx,url,times,booktabs, tabularx}
\usepackage[breaklinks=true]{hyperref}
\usepackage{cleveref}[2012/02/15]

\crefformat{footnote}{#2\footnotemark[#1]#3}

\newcommand{\var}[1]{{\operatorname{#1}}}

\title{Description and analysis of novelties introduced in DCASE Task 4 2022 \\  on the baseline system}

\name{Francesca Ronchini$^{1*}$,
       Samuele Cornell$^{2*}$,\thanks{$^*$ denotes equal contribution.}
       Romain Serizel$^{1}$, 
       Nicolas Turpault$^{1}$,
       }
\secondlinename{	  
       Eduardo Fonseca$^{3}$,
       Daniel P. W. Ellis$^{3}$
       }
 \address{$^1$ Universite de Lorraine, CNRS, Inria, Loria, Nancy, France \\     
         $^2$Department of Information Engineering, Universita Politecnica delle Marche, Italy \\
         $^3$ Google Research, United States
  }

\begin{document}

\ninept
\maketitle

\begin{sloppy}
\begin{abstract}
The aim of the Detection and Classification of Acoustic Scenes and Events Challenge Task 4 is to evaluate systems for the detection of sound events in domestic environments using an heterogeneous dataset.
The systems need to be able to correctly detect the sound events present in a recorded audio clip, as well as localize the events in time. This year's task is a follow-up of DCASE 2021 Task 4, with some important novelties. The goal of this paper is to describe and motivate these new additions, and report an analysis of their impact on the baseline system. 
We introduced three main novelties: the use of external datasets, including recently released strongly annotated clips from Audioset, the possibility of leveraging pre-trained models, and a new energy consumption metric to raise awareness about the ecological impact of training sound events detectors.  
The results on the baseline system show that leveraging open-source pre-trained on AudioSet improves the results significantly in terms of event classification but not in terms of event segmentation.
\end{abstract}


\section{Introduction}
\label{sec:intro}
Machine Listening has a substantial impact on applications such as noise monitoring in smart cities ~\cite{ciaburro2020improving, zinemanas2019mavd}, smart homes and home security solutions ~\cite{serizel2018large, ebbers2021self}, health monitoring systems ~\cite{alsina2017homesound}, bio-acoustics ~\cite{lostanlen2019robust}, and hearing aids ~\cite{huwel2020hearing}, among others. 
Sound Event Detection (SED) is one particularly important task in the field of Machine Listening. Its goal is to correctly output the class of different sound events present in an audio clip, together with each sound event's time boundaries ~\cite{mesaros2021sound}. Multiple events can be active simultaneously in each audio recording, but target sound events can also overlap with other non-target sound events. This research area is in continual development, attracting an expanding community. 

Since 2018, the Detection and Classification Acoustic Scene and Events (DCASE) Challenge Task 4 evaluates systems for detection of sound events in audio clips recorded in domestic environments. The challenge's main goal is to explore to what extent it is possible to exploit an unbalanced and heterogeneous dataset, e.g. containing a limited subset of weakly labeled and strongly labeled data and a larger amount of unlabeled data.
This is especially important as strongly labeled data is expensive and time-consuming to obtain while weakly labeled and in particular unlabeled data are much more accessible and scalable. 
Moreover, unlabeled data and weakly labeled data raise significantly fewer privacy concerns. In fact, weak labels could be obtained in an automated manner and, as such, human annotation is not required. This allows using on-device federated techniques which are more privacy-friendly.

Unlabeled data is usually leveraged via self-supervised learning. Indeed, past iterations of DCASE Challenge Task 4 have mainly explored this path, with most of the participants employing the mean-teacher technique \cite{meanteacher} for this purpose. 
However, another possible direction is leveraging pre-trained models from related tasks such as Sound Event Classification (SEC) to obtain a SED system, e.g. via pseudo-labeling, fine-tuning or by using the pre-trained models' internal activations as additional high-level features.
In the last couple of years several SEC models \cite{kong2020panns, gong2021ast} trained on the large AudioSet \cite{Gemmeke2017} have been open-sourced, making this approach particularly appealing.
The use of such pre-trained models has been recently boosted by advancements in self-supervised learning, which enables the training in an unsupervised way on massive amounts of unlabeled data. These models can then be fine-tuned for multiple downstream tasks. 
A shining example in the audio domain is Wav2Vec 2.0 \cite{baevski2020wav2vec}.
While powerful, these models are usually expensive to train and run (in particular self-attention based ones such as Wav2Vec 2.0 \cite{baevski2020wav2vec}), and unsuitable for widespread deployment on on-edge devices. 
On this premise, this year we introduced the following novelties for the DCASE Challenge Task 4:

\begin{itemize}

    \item we allowed the use of external datasets and embeddings extracted from open-source pre-trained models. Participants were encouraged to propose open-source models to use and external data sources, such as AudioSet \cite{Gemmeke2017}, which might also include real-word strongly annotations~\cite{hershey2021benefit}. 
    
    \item we developed a new codebase\footnote{Available at \href{https://github.com/DCASE-REPO/DESED_task/blob/master/recipes/dcase2022_task4_baseline}{\nolinkurl{github.com/DCASE-REPO/DESED\_task/blob/master/recipes/dcase2022\_task4\_baseline}}} and new baseline models to encourage participants to explore pre-trained models and such new datasets. 
    We present the results of such baselines in Section \ref{sec:impact}.
    
    \item finally, we introduced a new energy consumption metric based on CodeCarbon toolkit \cite{codecarbon}. This direction aims to foster interest among participants in finding new solutions for effective but also efficient SED systems. 
    This was mainly a pilot experiment as the reliability of CodeCarbon across various computational platforms has not been fully investigated. We explain in detail the new metric in Section \ref{sec:codecarbon}, together with the challenges that needs to be addressed in order to meaningfully compare the energy consumption of different models on different hardware platforms. 
\end{itemize}

The goal of this paper is to describe and motivate these new directions of this year's challenge, and to analyse their impact on the baseline performance.

\section{DCASE 2022 Challenge Task 4: Novelties and Motivations}
\label{sec:novelties}
In every edition, the DCASE Challenge Task 4 proposes to focus on new research questions, targeting specific aspects of the SED task that are considered interesting and timely for the research community, with the ultimate goal of advancing the state-of-the-art.

This year, we investigate three main aspects that are motivated by the availability of large scale datasets related to the SED task, the recent popularity of pre-trained generic audio representation and the growing concerns regarding the environmental impact of our digital life. These developments were proposed in order to target three key scientific questions related to SED systems that we believe are worth addressing.

\subsection{What is the impact of using external data and pre-trained models on SED systems?} 
\label{ssec:ext_data_question}

In the DCASE 2021 Challenge Task 4 \cite{dcase2021Task4}, we found that using a separation model trained in an unsupervised way using MixIT \cite{wisdom2020unsupervised} on the massive YFC100m dataset \cite{thomee2016yfcc100m} yielded a significant performance boost when used in conjunction with the baseline SED model on the evaluation set even if results on the development set were not that promising.  
Motivated by such result, this year we allowed participants to also use external data for the purpose of improving SED performance. 
To encourage participants to explore this path, we also provided a baseline which employs embeddings from two state-of-the-art popular models trained on AudioSet \cite{Gemmeke2017}: PANNs \cite{kong2020panns} and AST \cite{gong2021ast}.
In addition to being allowed to use AudioSet and these two models, participants were granted the use of other pre-trained models such as YAMNet \cite{hershey2017cnn} and also datasets not strictly related to SED such as MUSAN \cite{snyder2015musan} and ImageNet (used in AST \cite{gong2021ast} for example).
Participants were also welcomed to propose other pre-trained models and external datasets.
The full list of external resources allowed for the task can be found on the task website \footnote{\label{note1}\href{https://dcase.community/challenge2022/task-sound-event-detection-in-domestic-environments\#external-data-resources}{\nolinkurl{dcase.community/challenge2022/task-sound-event-detection-in-domestic-environments\#external-data-resources}}} and in \cite{Turpault2019_DCASE, Serizel2020_ICASSP}.

Each team was allowed to submit four different systems. 
However, in order to highlight the impact of external resources on SED systems, we required each team to submit at least one system that was not using external data. 

\subsection{Is strongly annotated real-world data necessary to build an effective SED system?}
\label{ssec:secondq}
In Hershey et al.~\cite{hershey2021benefit} it was found that strong labels (with temporally precise onset and offset together with sound-event class label) can bring substantial benefits in SEC applications even when they are provided for a small fraction of the total data, which can remain weakly labeled (only the sound class label is provided without any temporal precision). 
This is especially remarkable since manual annotating data is costly and time-consuming, but also bias-prone due to human errors and
disagreement on the perception of some sound event onsets and offsets. On the other hand, synthetic data is cheap to obtain, but also inherently mismatched with respect to real-world data, leading to potentially sub-optimal performance~\cite{ronchini2022benchmark}. 
In an attempt to mitigate this mismatch, and make the synthetic audio more realistic, in the past edition, non-target events have been included in the synthetic split of the training dataset \cite{ronchini2021impact}. 
This year, among the allowable external data source,  we also included 3470 strongly labeled recorded audio clips coming from AudioSet~\cite{hershey2021benefit}. The goal is to assess if the substantial improvement observed for SEC in Hershey et al.~\cite{hershey2021benefit} also translates to SED and to what extent it may be worth spending extra resources for more manual annotation. 

\subsection{What is the environmental footprint of our SED systems?}\label{sec:env_footprint}

Current state-of-the-art SED systems heavily rely on deep learning.
Numerous recent works \cite{henderson2020towards, parcollet2021energy} have raised concerns about the massive environmental costs of training deep learning models with large amount of parameters on massive amounts of data. 
For example, focusing on the audio domain, in Parcollet et al.~\cite{parcollet2021energy}, a study on the carbon footprint for ASR training was performed using the CodeCarbon toolkit \footnote{\href{https://codecarbon.io/}{https://codecarbon.io/}}, a software package that estimates the amount of energy consumption and carbon dioxide produced by the cloud or personal computing resources used to execute the code.
One of the key takeaways from this work is the tremendous inefficiency of many current ASR state-of-the-art models that trade off significant energy consumption (and thus pollution) for a marginal increase in performance that is likely not significant in actual deployment scenarios. 
These results raise important questions regarding the direction of ASR research, and suggests that the blind pursuit of the best possible performance in spite of the energy efficiency is likely not worth from a practical standpoint. 

Motivated by this study, this year we added support in the DCASE Challenge 2022 Task 4 baseline and repository for energy consumption benchmarking based on CodeCarbon. 
Participants were encouraged to submit their CodeCarbon estimated kilowatt-hour kWh energy consumption figures for each submission, both for model training and, contrary to Parcollet et al.~\cite{parcollet2021energy}, also inference on evaluation data. 
In fact, we argue that the energy consumption in inference is more important than the training one as the model could end up being deployed on thousands of devices and ran for years, with cumulative energy consumption quickly outpacing training phase.  
We describe more in detail the new CodeCarbon energy consumption metric and report energy consumption figures for this year baselines in Section \ref{sec:codecarbon}.

\section{DCASE 2022 Challenge Task 4 Baseline System} 
\label{sec:history}

This year's challenge baseline is based on a convolutional recurrent neural network (CRNN) but includes the main novelty of having the possibility of using features extracted from pre-trained models. CRNN was already found to be 
the architecture, mainly taken from \cite{jiakai2018mean}, is composed of a CNN module  followed by a 2-layers bi-directional gated recurrent unit (biGRU). The CNN has 7 layers, each composed of batch normalization, gated linear unit and dropout. 
Input features are Log-Mel Filterbank Energies extracted with a 128 ms window and 16 ms stride. 
The model is trained with the mean-teacher strategy \cite{meanteacher, jiakai2018mean} on audio data resampled at 16\,kHz and outputs one frame-wise prediction each 64 ms. 
To leverage more effectively weakly and unlabeled labeled data, attention pooling is employed, as outlined in \cite{jiakai2018mean}, to derive clip-level predictions from frame-level predictions.
From 2020, small changes, such as MixUp \cite{zhang2017mixup}, and hyper-parameters improvements have been implemented on the baseline system, based on top-ranked systems submitted every years from participants. 

\subsection{Integration with Pre-Trained Models}
\label{sec:embeddings}

As mentioned in Section \ref{ssec:ext_data_question}, among the main novelties in this year challenge is the allowance of pre-trained models and external data. Participants were encouraged to explore this direction by an expanded codebase in the official challenge repository. We implemented support for two different pre-trained models: PANNs and AST, which we use to extract embeddings to aid in the SED task. 
In detail, these embeddings are late-fused into the CRNN classifier described previously, before the biGRU module, by combining them with the features extracted by the CNN module. During training the pre-trained model is kept frozen, and it is used only as an embeddings extractor. The CRNN SED classifier is instead trained with the mean-teacher strategy as outlined before.   

We consider two types of embeddings extracted from pre-trained models internal activations: global embeddings, which are extracted at clip-level $\in \mathbb{R}^{C}$, and frame-wise embeddings, which instead are frame-wise $\in \mathbb{R}^{S_e \times C}$, where $C$ is the dimension of each vector and $S_e$ is the length of the embeddings sequence.

Regarding PANNs, global embeddings are extracted after the mean and max pooling layers, following the original work \cite{kong2020panns}. 
The frame-wise embeddings instead are extracted from the third convolutional block, after dropout. 
Regarding AST, also in this case we follow the original work for global embeddings \cite{gong2021ast} and derive them from the layer before the classification head. Frame-wise embeddings instead come from the last transformer layer. 

As illustrated in Figure \ref{fig:fusion}, left panel, global embeddings are simply fused with CNN features $\in \mathbb{R}^{S_c \times D}$ via concatenation at each time-step $s\in [1, \hdots S_c]$. In this work $D=128$. 
We firstly map the embeddings to same dimensionality as the CNN features $D$ and apply layer-normalization. Then these are concatenated with the CNN features at each time-step along the channel dimension, obtaining a tensor  $\mathbb{R}^{S_c \times 2D}$, another dense layer plus layer-normalization is used to shrink back the representation to $\mathbb{R}^{S_c \times D}$ before the biGRU. 
For frame-wise embeddings they cannot be concatenated right away as usually $S_e \neq S_c$, since PANNs, AST and the CNN front-end have different pooling factors. Thus we employ a single-layer BiGRU to encode the frame-wise embeddings into a representation with fixed dimensionality.
As illustrated in Figure \ref{fig:fusion}, right panel, we use the biGRU encoder last step output $\in \mathbb{R}^{H}$, where $H$ is 1024 channels,  twice the hidden size of the biGRU encoder, here we use 512 neurons. 
This encoded representation is then fed to the same pipeline employed for global embeddings.

\begin{figure}[ht]
\centering
\includegraphics[width=0.4\textwidth]{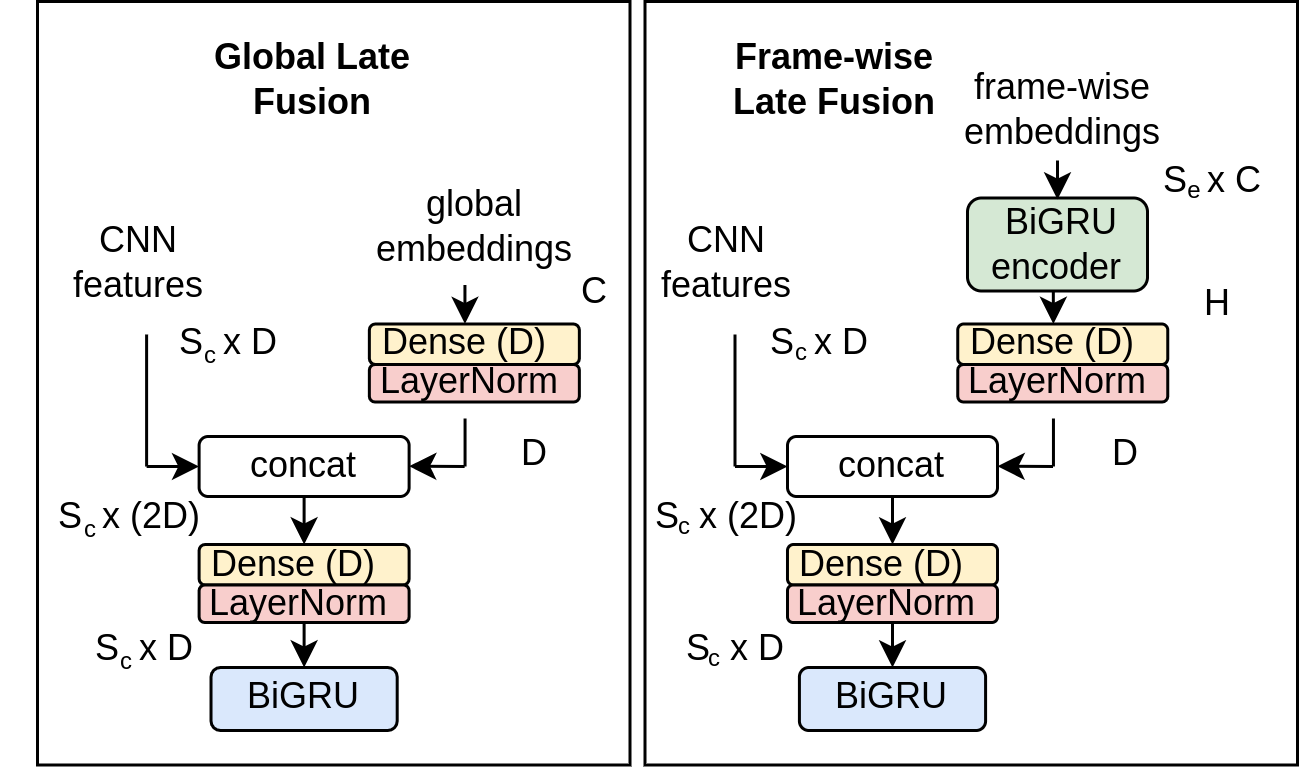}

\caption{\footnotesize{Combining pre-trained models embeddings with the CRNN Baseline via late-fusion before the biGRU module. Left: fusion with global embeddings. Right: fusion with frame-wise embeddings. We report the tensor dimensions as \textit{sequenceLength} $\times$ \textit{Channels}.}}
\label{fig:fusion}
\end{figure}

\subsection{CodeCarbon metric}
\label{sec:codecarbon}

As mentioned in Section \ref{ssec:ext_data_question},  this year we proposed to optionally report the energy consumption in kWh at training and test time. The goal here is to raise awareness regarding the environmental footprint of machine listening systems and SED systems in particular. 
Since every participant uses a different hardware platform to perform training and inference, absolute energy consumption figures are not directly comparable and cannot be used to assess each system efficiency.
To allow for a more fair and meaningful comparison we asked for the participants to also provide, for the dev-test and eval datasets, the energy consumption obtained when inference is performed with the CRNN baseline system. 
The CRNN baseline kWh can then be used effectively as a common measure, mitigating the factors of variations from hardware difference. 
As such, in order to analyze the systems performance in terms of SED together with their energy consumption, we used a tentative, trivial  energy weighted polyphonic sound detection score (EW-PSDS):
\begin{equation}
\mathrm{\var{EW-PSDS}} = \mathrm{PSDS} * \frac{\mathrm{kWh}_{\mathrm{baseline}}}{\mathrm{kWh}_{\mathrm{submission}}} 
\end{equation}

where PSDS is the polyphonic sound event detection scores \cite{bilen2020framework}, $\mathrm{kWh}_{\mathrm{baseline}}$ is the energy consumption reported for the baseline, and $\mathrm{kWh}_{\mathrm{submission}}$ is the energy consumption of the submitted system. 
Since providing energy consumption was not mandatory for participants, this initial experiment aims to provide insights to design more reliable protocols to obtain energy consumption report from challenge participants and more reliable metrics to report SED performance and energy consumption altogether. In particular, the proposed metric is very challenge-centric as it systematically relies on the energy consumption of the baseline as reference. Additionally, early results have shown that it is heavily biased by large energy consumption differences.

\section{Challenge Datasets and Evaluation Metrics}
\label{sec:dataset}

The dataset considered on this paper is the DESED dataset\footnote{\href{https://project.inria.fr/desed/}{https://project.inria.fr/desed/}} \cite{serizel:hal-02355573, turpault:hal-02160855}, which is the same as provided for the DCASE 2021 Challenge Task 4. 
It is composed of 10 seconds length audio clips either recorded in a domestic environment or synthesized to reproduce such an environment\footnote{For a detailed description of the DESED dataset and how it is generated the reader is referred to the original DESED article~\cite{turpault:hal-02160855} and DCASE 2021 task 4 webpage: \url{http://dcase.community/challenge2021}}.   
The synthetic part of the dataset is generated with Scaper \cite{salamon2017scaper}, a Python library for soundscape synthesis and augmentation. 

The foreground events (both target and non-target) are obtained from the Freesound Dataset (FSD50k) \cite{fonseca2020fsd50k}, while the background sounds are obtained from the SINS dataset (activity class “other”) \cite{dekkers2017sins} and TUT scenes 2016 development dataset \cite{mesaros2016tut}. The event co-occurences are computed on a set of strong annotations from Audioset~\cite{hershey2021benefit}.
More information regarding the generation of the DESED dataset can be found in Ronchini et al.~\cite{ronchini2021impact}.

As evaluation metrics we mainly use polyphonic sound event detection scores (PSDS) \cite{bilen2020framework} and consider two different applications scenarios. 
The first scenario targets the need of the systems to accurately detect the onset and offset of the sound event, while the second scenario penalizes more confusion between classes, but the temporal localization is less crucial \cite{ronchini2022benchmark}.

\begin{table}[t]
\centering
\footnotesize
\setlength{\tabcolsep}{4pt}
 \begin{tabular}{l|c|c|c|c}
   &  \multicolumn{2}{c|}{PSDS-1 $\uparrow$} & \multicolumn{2}{c}{PSDS-2 $\uparrow$}  \\
  \toprule
  & dev-test & eval & dev-test & eval \\
  Baseline & 0.336 & 0.315 & 0.536 & 0.543 \\
  \hline 
  w. AudioSet strong & 0.351 & \textbf{0.345} & 0.552 & 0.540\\
  \hline
  w. AST-frame & 0.313 & 0.290 & \textbf{0.722} & \textbf{0.678} \\
  w. AST-global & 0.205 & 0.192
 & 0.369 & 0.305 \\
  w. PANNs-frame & 0.354 & 0.304 &  0.635  &  0.597 \\
  w. PANNs-global & \textbf{0.375} &  0.308 & 0.668 & 0.584  \\
 \bottomrule
 \end{tabular}
 \caption{Results for the baseline system with additional external data or pre-trained models embeddings. We report PSDS for two application scenarios as described in Section \ref{sec:dataset}.}
 \label{tab:results}
\end{table}

\begin{table}[t!]
\centering
\footnotesize
\setlength{\tabcolsep}{2pt}
 \begin{tabular}{l|c|c|c|c|c|c}
   & 	\multicolumn{2}{c|}{{kWh $\downarrow$}} & \multicolumn{2}{c|}{EW-PSDS-1 $\uparrow$} & \multicolumn{2}{c|}{EW-PSDS-2 $\uparrow$} \\
     \toprule
     & Dev-test  & Eval &  Dev-test & Eval & Dev-test & Eval  \\
 \hline
  Baseline  & \textbf{0.030} & \textbf{0.617} & \textbf{0.336} & \textbf{0.315} & \textbf{0.536}  & \textbf{0.543} \\
  
  \hline
  w. AST-frame & 0.061 & 0.901 & 0.149 & 0.198 & 0.344 &  0.464 \\
  w. AST-global  &  0.063 & 0.873 & 0.097 & 0.136   &0.181 &  0.215 \\
  w. PANNs-frame  &  0.045 & 0.713 & 0.236 & 0.263 & 0.423 & 0.516 \\
  w. PANNs-global &  0.045 & 0.724 &  0.262  & 0.285 & 0.445 &  0.497 \\
 \bottomrule
 \end{tabular}
 \caption{Energy consumption (kWh) and Energy Weighted PSDS scores obtained on an Nvidia A100 GPU during inference on dev-test and evaluation.}
 \label{tab:energy}
\end{table}

\section{Impact of 2022 Challenge novelties on baseline system}
\label{sec:impact}
Tables \ref{tab:results} and \ref{tab:energy} report the results of this study. In Table \ref{tab:results} we study the impact of pre-trained models and additional strongly labeled data on the challenge baseline SED system described in Section \ref{sec:history}. 
Firstly, we can observe that the addition of strongly-labeled data from Audioset, as expected, improves the PSDS-1 but has practically no effect on PSDS-2. 
Instead, the use of embeddings from pre-trained models seems to have the opposite trend, it bring benefits mainly to PSDS-2, with AST-frame and PANNs-global coming on top. This is intuitive as PANNs and AST are trained to perform SEC. 
AST-global seems to perform very poorly. 
This is explained by the fact that the embeddings in this model are taken directly before the final classification output projection and thus the representation may be too much biased towards SEC.
On the other hand, as far as PANNs embeddings are concerned, global embeddings offer the best performance overall but the two are very close. Note that differently from AST, the global embeddings in PANNs are not taken from the layer just before the final linear classification projection but rather after the mean pooling layer.

The takeaway from these experiments is that downstream performance can heavily depend on the choice of the layer at which the pre-trained model embeddings are extracted. 
Instead of considering just one layer, future works could consider fusing features from many different layers, for example via self-attention. 

In Table \ref{tab:energy} we compare the energy consumption of the different architectural variations of the baseline system: PANNs versus AST and frame versus global embeddings. 
In detail we report kWh using CodeCarbon as explained in Section \ref{sec:codecarbon} on dev-test and evaluation. We also report the relative EW-PSDS scores so that we can take into account each model efficiency. 
We can notice that AST models consistently require more energy in inference than PANNs. This is in accordance with the number of parameters 80M for PANNs and 88.1M for AST, and with the fact that AST are based on self-attention which is more computationally expensive than convolution. 
Secondly we can notice that there is little to no difference between frame and global embeddings. This is unexpected because we know that the baseline with frame-wise embeddings should be slightly more computational demanding due to the biGRU encoder as illustrated in Figure \ref{fig:fusion}. This is likely due to the fact that the pre-trained model embedding extraction likely lead the energy consumption figure and CodeCarbon is simply not sensitive enough to pick up this subtle architectural difference. 
A more fair measure would be floating points operations (FLOP) for each frame prediction in output, however this has also the drawback that it is not easy to derive in a scalable way for multiple different models as current open source toolkits lack support for many operations. 

Regarding the energy-weighted PSDS measures we can see that overall the most efficient system seems to be the ``plain" baseline. This is because the pre-trained models used are much more computational intensive than the baseline, with one order of magnitude more parameters. More sophisticated techniques such as distillation could however mitigate this and drive down such energy consumption in inference. 
On the other hand, the EW-PSDS definition in Section \ref{sec:codecarbon} may in fact be too aggressive and too penalizing regarding energy consumption.
Nevertheless indeed the simple CRNN baseline still performs fairly good, it is remarkable that, for example in Table \ref{tab:results}, the best results for PSDS-1 are obtained without any additional pretrained model.

\section{CONCLUSION}
\label{sec:conclusions}
In this paper we presented the developments introduced in this year's DCASE Challenge Task 4 and analyzed their impact on the baseline performance. We focused on the possibility to use external datasets, in particular AudioSet with strong annotations, embeddings extracted with pre-trained models and on monitoring the energy consumption of the different systems at inference time. We have shown that using recorded (in-domain) clips from AudioSet with strong annotations together with synthetic soundscapes improves the PSDS-1 performance which focuses on accurately localizing the sound events in time. Using embeddings extracted from pre-trained SEC models (like PANNs or AST) improves the performance in terms of PSDS-2 which focuses on the accurate estimation on the sound event class (with loose constraints on the time localization). Yet, there are still some open questions regarding the possibility to efficiently exploit these external resources. Indeed, the baseline remains competitive with the systems using external data or pre-trained models while being substantially simpler. This aspect also reflect in terms of energy consumption of the different systems. In particular, the EW-PSDS of the models with pre-trained models is consistently lower than that of the baseline. Finally, the way monitoring energy consumption was introduced here remains naive and should be consolidated but the preliminary results open the way to considering SED systems under another angle.

\bibliographystyle{IEEEtran}
\bibliography{refs}

\begin{thebibliography}{10}
\providecommand{\url}[1]{#1}
\def\UrlFont{\rmfamily}
\providecommand{\newblock}{\relax}
\providecommand{\bibinfo}[2]{#2}
\providecommand\BIBentrySTDinterwordspacing{\spaceskip=0pt\relax}
\providecommand\BIBentryALTinterwordstretchfactor{4}
\providecommand\BIBentryALTinterwordspacing{\spaceskip=\fontdimen2\font plus
\BIBentryALTinterwordstretchfactor\fontdimen3\font minus
  \fontdimen4\font\relax}
\providecommand\BIBforeignlanguage[2]{{%
\expandafter\ifx\csname l@#1\endcsname\relax
\typeout{** WARNING: IEEEtran.bst: No hyphenation pattern has been}%
\typeout{** loaded for the language `#1'. Using the pattern for}%
\typeout{** the default language instead.}%
\else
\language=\csname l@#1\endcsname
\fi
#2}}

\bibitem{ciaburro2020improving}
G.~Ciaburro and G.~Iannace, ``Improving smart cities safety using sound events
  detection based on deep neural network algorithms,'' in \emph{Informatics},
  2020.

\bibitem{zinemanas2019mavd}
P.~Zinemanas, P.~Cancela, and M.~Rocamora, ``Mavd: A dataset for sound event
  detection in urban environments.'' \emph{DCASE Workshop}, 2019.

\bibitem{serizel2018large}
R.~Serizel, N.~Turpault, H.~Eghbal-Zadeh, and A.~P. Shah, ``Large-scale weakly
  labeled semi-supervised sound event detection in domestic environments,''
  \emph{arXiv preprint arXiv:1807.10501}, 2018.

\bibitem{ebbers2021self}
J.~Ebbers and R.~Haeb-Umbach, ``Self-trained audio tagging and sound event
  detection in domestic environments,'' in \emph{DCASE Workshop}, 2021.

\bibitem{alsina2017homesound}
R.~M. Alsina-Pag{\`e}s, J.~Navarro, F.~Al{\'\i}as, and M.~Herv{\'a}s,
  ``homesound: Real-time audio event detection based on high performance
  computing for behaviour and surveillance remote monitoring,'' \emph{Sensors},
  2017.

\bibitem{lostanlen2019robust}
V.~Lostanlen, J.~Salamon, and A.~e.~a. Farnsworth, ``Robust sound event
  detection in bioacoustic sensor networks,'' \emph{PloS one}, 2019.

\bibitem{huwel2020hearing}
A.~H{\"u}wel, K.~Adilo{\u{g}}lu, and J.-H. Bach, ``Hearing aid research data
  set for acoustic environment recognition,'' in \emph{Proc. of ICASSP}, 2020.

\bibitem{mesaros2021sound}
A.~Mesaros, T.~Heittola, T.~Virtanen, and M.~D. Plumbley, ``Sound event
  detection: A tutorial,'' \emph{IEEE Signal Processing Magazine}, 2021.

\bibitem{meanteacher}
A.~Tarvainen and H.~Valpola, ``Mean teachers are better role models:
  Weight-averaged consistency targets improve semi-supervised deep learning
  results,'' in \emph{Advances in neural information processing systems}, 2017.

\bibitem{kong2020panns}
Q.~Kong, Y.~Cao, T.~Iqbal, and Y.~e.~a. Wang, ``Panns: Large-scale pretrained
  audio neural networks for audio pattern recognition,'' \emph{IEEE TASLP},
  2020.

\bibitem{gong2021ast}
Y.~Gong, Y.-A. Chung, and J.~Glass, ``Ast: Audio spectrogram transformer,''
  \emph{arXiv preprint arXiv:2104.01778}, 2021.

\bibitem{Gemmeke2017}
J.~F. Gemmeke, D.~P.~W. Ellis, and D.~e.~a. Freedman, ``Audio set: An ontology
  and human-labeled dataset for audio events,'' in \emph{Proc. of ICASSP},
  2017.

\bibitem{baevski2020wav2vec}
A.~Baevski, Y.~Zhou, A.~Mohamed, and M.~Auli, ``wav2vec 2.0: A framework for
  self-supervised learning of speech representations,'' \emph{Advances in
  Neural Information Processing Systems}, 2020.

\bibitem{hershey2021benefit}
S.~Hershey, D.~P. Ellis, and E.~e.~a. Fonseca, ``The benefit of
  temporally-strong labels in audio event classification,'' in \emph{Proc. of
  ICASSP}, 2021.

\bibitem{codecarbon}
V.~Schmidt, K.~Goyal, and A.~J. et~al., ``{CodeCarbon: Estimate and Track
  Carbon Emissions from Machine Learning Computing},'' 2021.

\bibitem{dcase2021Task4}
F.~Ronchini, S.~Cornell, and N.~e.~a. Turpault, ``{DCASE} 2021 {Task} 4
  {Challenge},'' \url{https://dcase.community/challenge2021}, 2021.

\bibitem{wisdom2020unsupervised}
S.~Wisdom, E.~Tzinis, and H.~e.~a. Erdogan, ``Unsupervised sound separation
  using mixture invariant training,'' \emph{Advances in Neural Information
  Processing Systems}, 2020.

\bibitem{thomee2016yfcc100m}
B.~Thomee, D.~A. Shamma, and G.~e.~a. Friedland, ``Yfcc100m: The new data in
  multimedia research,'' \emph{Communications of the ACM}, 2016.

\bibitem{hershey2017cnn}
S.~Hershey, S.~Chaudhuri, and D.~P. e.~a. Ellis, ``Cnn architectures for
  large-scale audio classification,'' in \emph{Proc. of ICASSP}, 2017.

\bibitem{snyder2015musan}
D.~Snyder, G.~Chen, and D.~Povey, ``Musan: A music, speech, and noise corpus,''
  \emph{arXiv preprint arXiv:1510.08484}, 2015.

\bibitem{Turpault2019_DCASE}
N.~Turpault, R.~Serizel, A.~Parag~Shah, and J.~Salamon, ``{Sound event
  detection in domestic environments with weakly labeled data and soundscape
  synthesis},'' in \emph{{DCASE Workshop}}, 2019.

\bibitem{Serizel2020_ICASSP}
R.~Serizel, N.~Turpault, A.~Shah, and J.~Salamon, ``{Sound event detection in
  synthetic domestic environments},'' in \emph{{Proc. of ICASSP }}, 2020.

\bibitem{ronchini2022benchmark}
F.~Ronchini and R.~Serizel, ``A benchmark of state-of-the-art sound event
  detection systems evaluated on synthetic soundscapes,'' in \emph{Proc. of
  ICASSP}, 2022.

\bibitem{ronchini2021impact}
F.~Ronchini, R.~Serizel, N.~Turpault, and S.~Cornell, ``The impact of
  non-target events in synthetic soundscapes for sound event detection,''
  \emph{arXiv preprint arXiv:2109.14061}, 2021.

\bibitem{henderson2020towards}
P.~Henderson, J.~Hu, and J.~a.~a. Romoff, ``Towards the systematic reporting of
  the energy and carbon footprints of machine learning,'' \emph{Journal of
  Machine Learning Research}, 2020.

\bibitem{parcollet2021energy}
T.~Parcollet and M.~Ravanelli, ``The energy and carbon footprint of training
  end-to-end speech recognizers,'' 2021.

\bibitem{jiakai2018mean}
L.~JiaKai, ``Mean teacher convolution system for dcase 2018 task 4,'' DCASE2018
  Challenge, Tech. Rep., 2018.

\bibitem{zhang2017mixup}
H.~Zhang, M.~Cisse, and Y.~N. e.~a. Dauphin, ``mixup: Beyond empirical risk
  minimization,'' \emph{arXiv preprint arXiv:1710.09412}, 2017.

\bibitem{bilen2020framework}
{\c{C}}.~Bilen, G.~Ferroni, and F.~e.~a. Tuveri, ``A framework for the robust
  evaluation of sound event detection,'' in \emph{Proc. of ICASSP}, 2020.

\bibitem{serizel:hal-02355573}
R.~Serizel, N.~Turpault, A.~Shah, and J.~Salamon, ``{Sound event detection in
  synthetic domestic environments},'' in \emph{Proc. of ICASSP}, 2020.

\bibitem{turpault:hal-02160855}
N.~Turpault, R.~Serizel, A.~Parag~Shah, and J.~Salamon, ``{Sound event
  detection in domestic environments with weakly labeled data and soundscape
  synthesis},'' in \emph{Detection and Classification of Acoustic Scenes and
  Events, Workshop, DCASE}, 2019.

\bibitem{salamon2017scaper}
J.~Salamon, D.~MacConnell, M.~Cartwright, P.~Li, and J.~P. Bello, ``Scaper: A
  library for soundscape synthesis and augmentation,'' in \emph{Proc. of
  WASPAA}, 2017.

\bibitem{fonseca2020fsd50k}
E.~Fonseca, X.~Favory, J.~Pons, and e.~a. Font, ``Fsd50k: an open dataset of
  human-labeled sound events,'' \emph{arXiv preprint arXiv:2010.00475}, 2020.

\bibitem{dekkers2017sins}
G.~Dekkers, S.~Lauwereins, and T.~et~al., ``The sins database for detection of
  daily activities in a home environment using an acoustic sensor network,'' in
  \emph{DCASE Workshop}, 2017.

\bibitem{mesaros2016tut}
A.~Mesaros, T.~Heittola, and T.~Virtanen, ``Tut database for acoustic scene
  classification and sound event detection,'' in \emph{EUSIPCO}, 2016.

\end{thebibliography}

%
%
%
%
%
%
%
%
%

\end{sloppy}
\end{document}